\documentstyle[graphicx,12pt]{article}
\begin{document}

\small
\hoffset=-1truecm
\voffset=-2truecm
\title{\bf The strong field gravitational lensing in the Schwarzschild black hole pierced by a cosmic string}
\author{Jingyun Man \hspace {1cm} Huawen Wang \hspace {1cm} Hongbo Cheng\footnote
{E-mail address:
hbcheng@ecust.edu.cn}\\
Department of Physics, East China University of Science and
Technology,\\ Shanghai 200237, China\\
The Shanghai Key Laboratory of Astrophysics,\\ Shanghai 200234,
China}

\date{}
\maketitle

\begin{abstract}
In this work the gravitational lensing in the strong field limit
around the Schwarzschild black hole pierced by a cosmic string is
studied. We find that the deflection angle and the time delay of
the relativistic images depend on the tension of cosmic string. It
is interesting that the deflection angle is greater when the
tension of cosmic string is stronger. The time delay between two
images is more obvious in the case of weaker tension.
\end{abstract}
\vspace{6cm} \hspace{1cm} PACS number(s): 95.30.Sf, 04.70.-s

\newpage

\noindent \textbf{1.\hspace{0.4cm}Introduction}

As an important application of general relativity the
gravitational lensing is on the deflection of electromagnetic
radiation in a gravitational field. Recently more efforts have
been contributed to the gravitational lensing and the issue has
been developed greatly [1-3]. In fact, it is difficult to
investigate the relations between the deflection angle and the
properties of gravitational source because the deflection angle of
light passing close to a compact and massive source is shown in
integral forms. The integral expressions can be dealt with in the
limiting cases such as weak field approximation and strong field
limit. When the light goes very close to a heavy compact body like
a black hole, an infinite series of images generate, which provide
more information about the nature of the black hole's surrounding,
and the process is thought as strong field limit. Under the strong
field limit the integral expression for deflection angle is
discussed around the radius of photon sphere which leads the
deflection angle to be infinity. In these cases the weak field
approximation is not valid and the strong gravitational lensing
can help us to explore the characteristics of the gravitational
source. More topics within the region of the strong field have
been discussed. We can list that the strong gravitational lensing
was treated in a Schwarzschild black hole [4, 5], gravitational
source with naked sigularities [6], a Reissner-Nordstrom black
hole [7], a GMGHS charged black hole [8], a spining black hole [9,
10], a braneworld black hole [11, 12], an Einstein-Born-Infeld
black hole [13], a black hole in Brans-Dicke theory [14], a black
hole with Barriola-Vilenkin monopole [15, 16] and the deformed
Horava-Lifshitz black hole [17], etc..

As a kind of topological defects formed during the phase
transition in the early universe due to the Kibble mechanism, the
cosmic strings have been a focus for many years [18]. According to
the inflationary models resulting form string theory the cosmic
string networks which consist of long strings and string loops
appeared at the end of inflation [19, 20]. The evolution of cosmic
string loops have been studied in various backgrounds [21-22]. The
cosmic string networks evolve then the cosmic strings interacted
with the other gravitational source or each other. The
Schwarzschild black holes pierced by cosmic strings could have
formed in the process of early phase transition and survive. It
was shown that the Schwarzschild black hole pierced by a cosmic
string with full $U(1)$ Abelian-Higgs model has long-range "hair",
excluding the no hair conjecture [23, 24]. A cosmic string
piercing a Schwarzschild black hole has also been considered in
the limit of thin string and the corrections were brought about in
the Schwarzschild solution [25]. In the new metric the angular
variable $\phi$ is replaced by $\gamma\phi$, where the parameter
$\gamma$ is associated with the deficit angle by
$\Delta=2\pi(1-\gamma)$ due to the cosmic string. In order to
explore the properties of the spacetime, the geodesic equation
near this kind of gravitational source is investigated carefully
[26].

It is interesting to probe the massive source in different
directions. The main purpose of this work is to investigate the
gravitational lensing on the Schwarzschild black hole pierced by a
cosmic string. We plan to study the deflection angle of
electromagnetic radiation passing very close to this kind of
massive source with cosmic string to show the angle deviation
resulting from the cosmic string by means of Bozza's device put
forward in Ref. [27]. We plot the dependence of the observational
gravitational lensing coefficients on the parameter subject to the
cosmic string going through the black hole. Furthermore we compute
the time delay between images generated by this kind of massive
source to show the time difference due to the properties of the
lens. Our results will be listed in the end.

\vspace{0.8cm} \noindent \textbf{2.\hspace{0.4cm}The deflection
angle of the Schwarzschild black hole pierced by a cosmic string}

The explicit form of the metric that describing a Schwarzschild
black hole pierced by a cosmic string is [26],

\begin{equation}
ds^{2}=A(r)dt^{2}-B(r)dr^{2}-r^{2}(d\theta^{2}+\gamma^{2}\sin^{2}\theta
d\varphi^{2})
\end{equation}

\noindent where

\begin{equation}
A(r)=B^{-1}(r)=1-\frac{2GM}{r}
\end{equation}

\begin{equation}
\gamma=1-4G\mu
\end{equation}

\noindent and $G$ is Newton's constant. $\mu$ is the tension of
string.

We discuss the gravitational lensing in the strong field limit
around the Schwarzschild black hole pierced by a cosmic string
according to the methods of [1, 27]. Here we choose that both the
observer and the gravitational source lie in the equatorial plane
with condition $\theta=\frac{\pi}{2}$ for simplicity. The whole
trajectory of the photon is subject to the same plane that is
perpendicular to the long cosmic string. On the hypersurface with
$\theta=\frac{\pi}{2}$ the metric reads,

\begin{equation}
ds^{2}=A(r)dt^{2}-B(r)dr^{2}-C(r)d\varphi^{2}
\end{equation}

\noindent where

\begin{equation}
C(r)=\gamma^{2}r^{2}
\end{equation}

\noindent The deflection angle for the electromagnetic radiation
emitted from the distant place depends on the closest distance
between the ray and the source and can be expressed as [28],

\begin{equation}
\alpha(r_{0})=I(r_{0})-\pi
\end{equation}

\noindent where

\begin{equation}
I(r_{0})=2\int_{r_{0}}^{\infty}\frac{\sqrt{B(r)}}{\sqrt{C(r)}
\sqrt{\frac{C(r)A(r_{0})}{C(r_{0})A(r)}}-1}dr
\end{equation}

\noindent and here $r_{0}$ represents the minimum distance from
the photon trajectory to the gravitational source. If the
denominator of expression (7) is equal to the zero, the deflection
angle will be divergent, then the photons will move around the
source forever instead of leaving. The null denominator in the
integral expression leads the equation
$\frac{A(r)}{A(r_{0})}=\frac{C(r)}{C(r_{0})}$ whose largest root
is named as the radius of the photon circle as mapping of photon
sphere in the equatorial plane [6, 28]. Having solved the
equation, we obtain the radius of the photon sphere in the metric
of Schwarzschild black hole pierced by a long cosmic string (4) as
follow,

\begin{equation}
r_{m}=3GM
\end{equation}

\noindent It is interesting that the photon sphere radius is the
same as that of Schwarzschild metric without cosmic string. The
cosmic string does not change the event horizon either. Of course
the radius of photon sphere is larger than the black hole radius.
It is clear that the cosmic string does not modify the photon
sphere radius and the event horizon within the plane that the
string is perpendicular to. With $C(r)=\gamma^{2}r^{2}$, we
re-derive to find that the following expressions consisting of
line elements of metric describe the deflection angle and
coefficients in the strong field limit and have the same form as
Bozza's in Ref. [27]. We can make use of Bozza's results and just
let $C(r)=\gamma^{2}r^{2}$ to explore our topic. Further we wonder
how the parameter $\gamma$ produced by cosmic string brings about
the influence on the gravitational lensing. According to what
mentioned above, we follow the procedure by Bozza in Ref. [27] to
express the deflection angle in the strongfield limit as follow,

\begin{equation}
\alpha(\theta)=-\bar{a}\ln(\frac{\theta
D_{OL}}{u_{m}}-1)+\bar{b}+O(u-u_{m})
\end{equation}

\noindent where $D_{OL}$ means the distance between observer and
gravitational lens. The impact parameter is [28],

\begin{equation}
u=\sqrt{\frac{C(r_{0})}{A(r_{0})}}
\end{equation}

\noindent Certainly,

\begin{equation}
u_{m}=u|_{r_{0}=r_{m}}
\end{equation}

\noindent The strong field limit coefficients $\bar{a}$ and
$\bar{b}$ are expressed as,

\begin{equation}
\bar{a}=\frac{R(0,r_{m})}{2\sqrt{d_{m}}}
\end{equation}

\begin{equation}
\bar{b}=-\pi+I_{R}(r_{m})+\bar{a}\ln\frac{2d_{m}}{A(r_{m})}
\end{equation}

\noindent where

\begin{equation}
d_{m}=d|_{r_{0}=r_{m}}
\end{equation}

\noindent It is necessary to explain the variables $\bar{a}$ and
$\bar{b}$. We define some functions as follow,

\begin{equation}
R(z,r_{0})=\frac{2\sqrt{B(r)A(r)}}{C(r)\frac{dz}{dr}}\sqrt{C(r_{0})}
\end{equation}

\begin{equation}
f(z,r_{0})=\frac{1}{\sqrt{A(r_{0})-[(1-A(r_{0}))z+A(r_{0})]\frac{C(r_{0})}{C(r)}}}
\end{equation}

\begin{equation}
f_{0}(z,r_{0})=\frac{1}{\sqrt{cz+dz^{2}}}
\end{equation}

\begin{equation}
I_{D}(r_{0})=\int_{0}^{1}R(0,r_{m})f_{0}(z,r_{0})dz
\end{equation}

\begin{equation}
I_{R}(r_{0})=\int_{0}^{1}[R(z,r_{0})f(z,r_{0})-R(0,r_{m})f_{0}(z,r_{0})]dz
\end{equation}

\noindent where

\begin{equation}
z=\frac{A(r)-A(r_{0})}{1-A(r_{0})}
\end{equation}

\begin{equation}
c=\frac{1-A(r_{0})}{C(r_{0})A'(r_{0})}(C'(r_{0})A(r_{0})-C(r_{0})A'(r_{0}))
\end{equation}

\begin{equation}
d=\frac{(1-A(r_{0}))^{2}}{2C^{2}(r_{0})A'^{3}(r_{0})}
[2C_{0}C'_{0}A'^{2}_{0}+(C_{0}C''_{0}-2C'^{2}_{0})A_{0}A'_{0}
-C_{0}C'_{0}A_{0}A''_{0}]
\end{equation}

\noindent with $A_{0}=A(r_{0})$, $C_{0}=C(r_{0})$. Now following
the procedure of Ref. [27, 28], we derive the expression of
deflection angle for the Schwarzschild black hole pierced by a
cosmic string under strong field limit. Substituting the metric
(4) into the equations above, we obtain that

\begin{equation}
R(z,r_{0})=\frac{2}{\gamma}
\end{equation}

\begin{equation}
f(z,r_{0})=[-\frac{2GM}{r_{0}}z^{3}-(1-\frac{6GM}{r_{0}})z^{2}
+2(1-\frac{3GM}{r_{0}})z]^{-\frac{1}{2}}
\end{equation}

\begin{equation}
f_{0}(z,r_{0})=[2(1-\frac{3GM}{r_{0}})z-(1-\frac{6GM}{r_{0}})z^{2}]^{-\frac{1}{2}}
\end{equation}

\noindent According to the defined functions and equations above,
the coefficients $\bar{a}$, $\bar{b}$ and the minimum impact
parameter of the deflection angle can be written as,

\begin{equation}
\bar{a}=\frac{1}{\gamma}=\frac{1}{1-4G\mu}
\end{equation}

\begin{equation}
\bar{b}=\frac{4}{1-4G\mu}\ln(3-\sqrt{3})+\frac{1}{1-4G\mu}\ln6-\pi
\end{equation}

\begin{equation}
u_{m}=3\sqrt{3}(1-4G\mu)GM
\end{equation}

\noindent The deflection angle of the Schwarzschild black hole
pierced by a cosmic string is,

\begin{equation}
\alpha(\theta)=-\frac{1}{1-4G\mu}\ln(\frac{\theta D_{OL}}
{3\sqrt{3}(1-4G\mu)GM}-1)+\frac{4}{1-4G\mu}\ln(3-\sqrt{3})
+\frac{1}{1-4G\mu}\ln6-\pi
\end{equation}

\noindent It is clear that the tension of cosmic string modifies
the deflection angle subject to the standard Schwarzschild black
hole without cosmic string. The deviation from the comic string
going through the massive source helps us to distinguish a pure
Schwarzschild black hole from that one involving a cosmic string.
The cosmic string tension $\mu$ appears in our results. Our
results will recover to those of standard Schwarzschild black hole
with $\mu=0$. We calculate these coefficients and observables
according to the equations above and plot their dependence on the
cosmic string tension in the figures. From Fig. 1 we show that the
two strong gravitational lensing coefficients $\bar{a}$ and
$\bar{b}$ both become larger with the increase of tension $\mu$.
They increase quickly as the tension is large enough. According to
Eq. (29), we also show the dependence of the deflection angle on
the corrections from the cosmic string in Fig. 2. Every curve
shows the effect from cosmic string on the deflection angle for a
definite value of $\frac{\theta D_{OL}}{GM}$. The negative value
of the angle $\alpha$ represents that the images locate on the
another side of the gravitational source. The curves of the
dependence of deflection angle on $G\mu$ with various values of
$\frac{\theta D_{OL}}{GM}$ respectively are similar. The
deflection angle becomes larger as the tension of cosmic string
increases. When the increasing tension of cosmic string is large
enough, the image will deviate more greatly. It should also be
emphasized that the influences from cosmic string on the
deflection angle $\alpha(\theta)$ are evident.

\vspace{0.8cm} \noindent \textbf{3.\hspace{0.4cm}The time delay
between images generated by the Schwarzschild black hole pierced
by a cosmic string}

This section is contributed to the expression of time delay
between two images subject to the massive source involving a
cosmic string. According to Ref. [28, 29], the time difference
between two photons travelling on different trajectories is
expressed as,

\begin{equation}
T_{1}-T_{2}=\widetilde{T}(r_{0,1})-\widetilde{T}(r_{0,2})
+2\int_{r_{0,1}}^{r_{0,2}}\frac{\widetilde{P_{1}}(r,r_{0,1})}{\sqrt{A(r_{0,1})}}
dr
\end{equation}

\noindent where

\begin{equation}
\widetilde{P_{1}}(r,r_{0})=\sqrt{\frac{B(r)A(r_{0})}{A(r)}}
\end{equation}

\noindent and the time duration for the light ray to wind around
the gravitational source is devoted in the strong field limit

\begin{equation}
\widetilde{T}(r_{0})=-\widetilde{a}\ln(\frac{u}{u_{m}}-1)+\widetilde{b}
+O(u-u_{m})
\end{equation}

\noindent with

\begin{equation}
\widetilde{a}=\frac{\widetilde{R(0, r_{m})}}{2\sqrt{d_{m}}}
\end{equation}

\begin{equation}
\widetilde{b}=-\pi+\widetilde{b}_{D}+\widetilde{b}_{R}
+\widetilde{a}\ln\frac{cr_{m}^{2}}{u_{m}}
\end{equation}

\begin{equation}
\widetilde{b}_{D}=2\widetilde{a}\ln\frac{2(1-A(r_{m}))}{A'(r_{m})r_{m}}
\end{equation}

\begin{equation}
\widetilde{b}_{R}=\int_{0}^{1}[\widetilde{R}(z, r_{m})f(z, r_{m})
-\widetilde{R}(0, r_{m})f_{0}(z, r_{m})]dz
\end{equation}

\begin{equation}
\widetilde{R}(z, r_{0})=\frac{2(1-A(r_{0}))}{A'(r)}
\widetilde{P}_{1}(r, r_{0})(1-\frac{1}{\sqrt{A(r_{0})}f(z,
r_{0})})
\end{equation}

\noindent while

\begin{equation}
u-u_{m}=\theta D_{OL}-u_{m}=\bar{c}(r_{0}-r_{m})^{2}
\end{equation}

\noindent We choose $\alpha(\theta)=2n\pi\pm\bar{\gamma}$, and
$\bar{\gamma}$ is the angular separation between the source and
the optical axis as seen from the lens. The position of the $n$-th
relativistic image can be approximated as,

\begin{equation}
u_{0,
n}=u_{m}[1+\exp(\frac{\bar{b}-2n\pi\pm\bar{\gamma}}{\gamma\bar{a}})]
\end{equation}

In the case that the two images lie on the same side of the lens,
their time delay is,

\begin{equation}
\Delta T_{n, m}^{s}=2\pi(n-m)\frac{\widetilde{a}}{\bar{a}}
+2\sqrt{\frac{B(r_{m})}{A(r_{m})}}\sqrt{\frac{u_{m}}{\bar{c}}}
e^{\frac{\bar{b}}{2\bar{a}}}[\exp(-\frac{2m\pi\mp\bar{\gamma}}{2\bar{a}})
-\exp(-\frac{2n\pi\mp\bar{\gamma}}{2\bar{a}})]
\end{equation}

\noindent where the minus sign means that the two images are on
the same side of the source and the positive one for the images
standing on the other side.

The time difference between the images on opposite side of the
lens is,

\begin{equation}
\Delta T_{n,
m}^{o}=[2\pi(n-m)-2\bar{\gamma}]\frac{\widetilde{a}}{\bar{a}}
+2\sqrt{\frac{B(r_{m})}{A(r_{m})}}\sqrt{\frac{u_{m}}{\bar{c}}}
e^{\frac{\bar{b}}{2\bar{a}}}[\exp(-\frac{2m\pi-\bar{\gamma}}{2\bar{a}})
-\exp(-\frac{2n\pi+\bar{\gamma}}{2\bar{a}})]
\end{equation}

\noindent We also connect the metric (4) with the expressions for
time delay to obtain the coefficients,

\begin{equation}
\widetilde{a}=6\sqrt{3}GM
\end{equation}

\begin{equation}
\bar{c}=\frac{3}{4}(\sqrt{3}-1)\frac{1-4GM}{GM}
\end{equation}

\noindent The time delay between two images for the same side case
is,

\begin{eqnarray}
\Delta T_{n, m}^{s}=12\sqrt{3}\pi(n-m)\pi(1-4G\mu)GM\hspace{5cm}\nonumber\\
+36\sqrt{2}(3-\sqrt{3})^{\frac{3}{2}}e^{-\frac{\pi}{2}(1-4G\mu)}
[e^{-m\pi(1-4G\mu)}-e^{-n\pi(1-4G\mu)}]GM
\end{eqnarray}

\noindent and here $\bar{\gamma}=0$.

We also obtain the time delay for the images lies on the opposite
side,

\begin{eqnarray}
\Delta T_{n, m}^{o}=6\sqrt{3}[2\pi(n-m)-2\bar{\gamma}](1-4G\mu)GM\hspace{5cm}\nonumber\\
+3\sqrt{2}(3-\sqrt{3})^{\frac{3}{2}}e^{-\frac{\pi}{2}(1-4G\mu)}
[e^{-\frac{2m\pi-\bar{\gamma}}{2}(1-4G\mu)}
-e^{-\frac{2m\pi+\bar{\gamma}}{2}(1-4G\mu)}]
\end{eqnarray}

\noindent It is significant that the time difference between two
images depends on the parameter belonging to the cosmic string,
which can also help us to detect the possibility that the massive
source has involved a cosmic string. In order to exhibit the
influence from the cosmic string on the time difference in the
strong field limit clearly. We show the dependence of the time
delay between the first and the second relativistic images lying
on the same side or opposite sides of the source on the tension of
cosmic string graphically in Fig. 3 and Fig. 4 respectively. The
two curves resemble each other. The time delay will become longer
when the string tension is stronger. We hope that the astronomical
observations for observables such as deflection angle, time delay
etc. will  become a new way to explore the cosmic string in
future.

\vspace{0.8cm} \noindent \textbf{3.\hspace{0.4cm}Summary}

In this work we research on the gravitational lensing in the
strong field limit around the Schwarzschild black hole pierced by
a cosmic string. It is interesting that the cosmic string leads
the minimum impact parameter $u_{m}$ to decrease although the
radius of photon circle has nothing to do with the string, which
means that the black hole has less time to attract photons
circling around it because of the deficit angle and it needs them
to be shot at smaller impact parameters in order to capture them.
We reveal that the strong gravitational lensing coefficients and
the deflection angle increase as the tension of cosmic string is
greater. We indicate that the stronger string tension also leads
the longer time delay between two relativistic images. The
deviation of the deflection angle and some other modifications
from cosmic string are manifest and distinct and certainly
different from those in some other cases such as the massive
source involving global monopole [16]. Observing these phenomena
supplies us a new way to explore the cosmic string. The cosmic
strings can produce some other distinct characteristics besides
two images presented in Ref. [18]. The related topics such as the
deflection angle of the light along the off-equator paths need to
be studied further in future.

\vspace{3cm}

\noindent\textbf{Acknowledgement}

This work is supported by NSFC No. 10875043 and is partly
supported by the Shanghai Research Foundation No. 07dz22020.

\newpage
\begin{figure}
\setlength{\belowcaptionskip}{10pt} \centering
  \includegraphics[width=15cm]{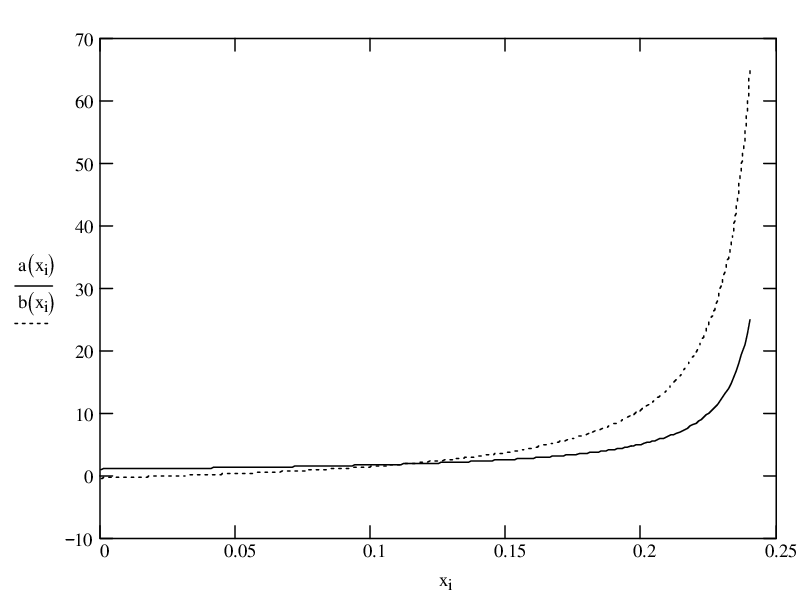}
  \caption{The solid and dot curves corresponds to the dependence of the coefficient of
  strong field limit $\bar{a}$ and $\bar{b}$ respectively on the parameter $x=G\mu$
  representing the tension of cosmic string piercing the Schwarzschild black hole.}
\end{figure}

\newpage
\begin{figure}
\setlength{\belowcaptionskip}{10pt} \centering
  \includegraphics[width=15cm]{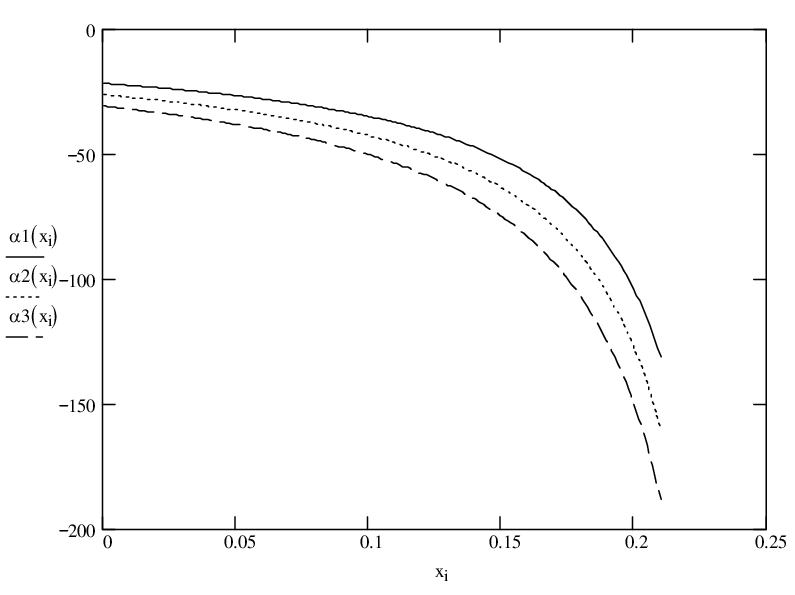}
  \caption{The solid, dot and dashed curves correspond to the dependence of the deflection angle $\alpha$ on the parameter $x=G\mu$
  representing the tension of cosmic string piercing the Schwarzschild black hole for $\frac{\theta D_{OL}}{GM}=10^{10}, 10^{12},
  10^{14}$ in natural units
  respectively.}
\end{figure}

\newpage
\begin{figure}
\setlength{\belowcaptionskip}{10pt} \centering
  \includegraphics[width=15cm]{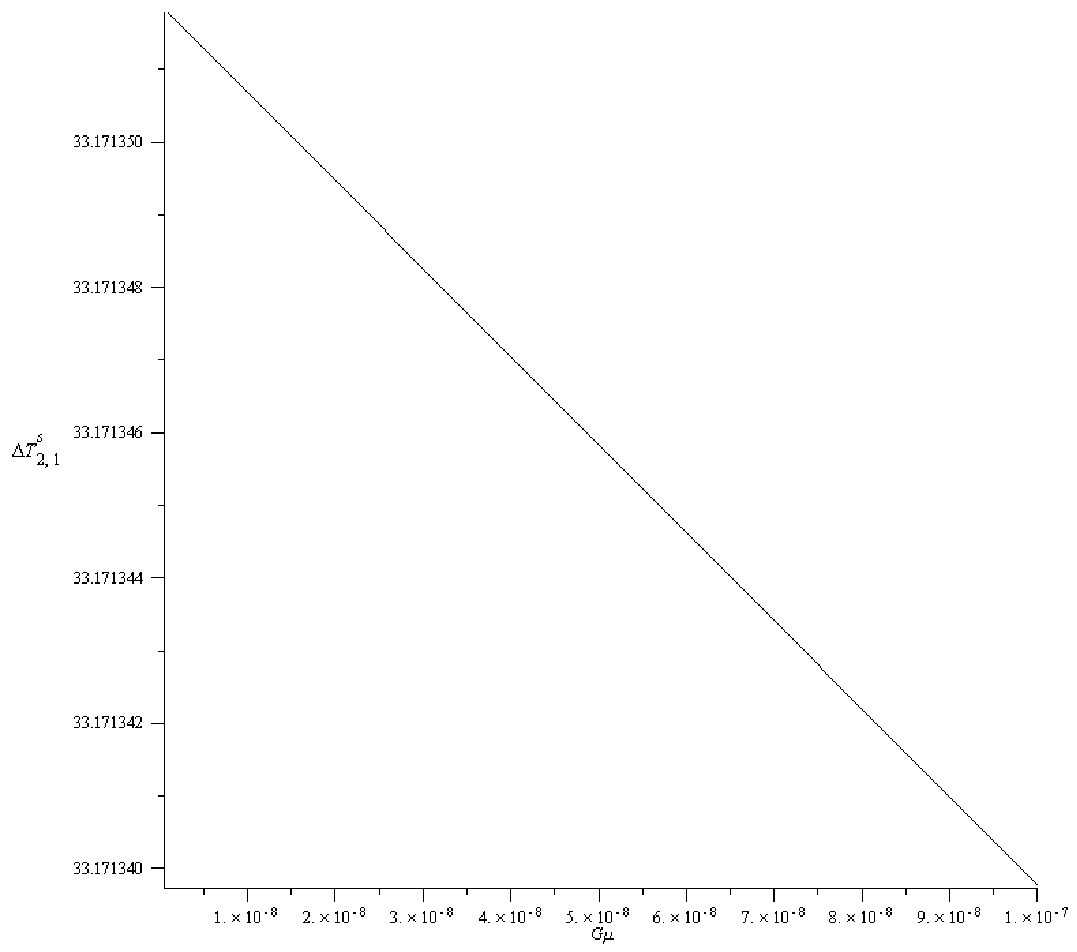}
  \caption{The dependence of time delay between two images on the same side on the parameter $G\mu$
  representing the tension of cosmic string piercing the Schwarzschild black hole.}
\end{figure}

\newpage
\begin{figure}
\setlength{\belowcaptionskip}{10pt} \centering
  \includegraphics[width=15cm]{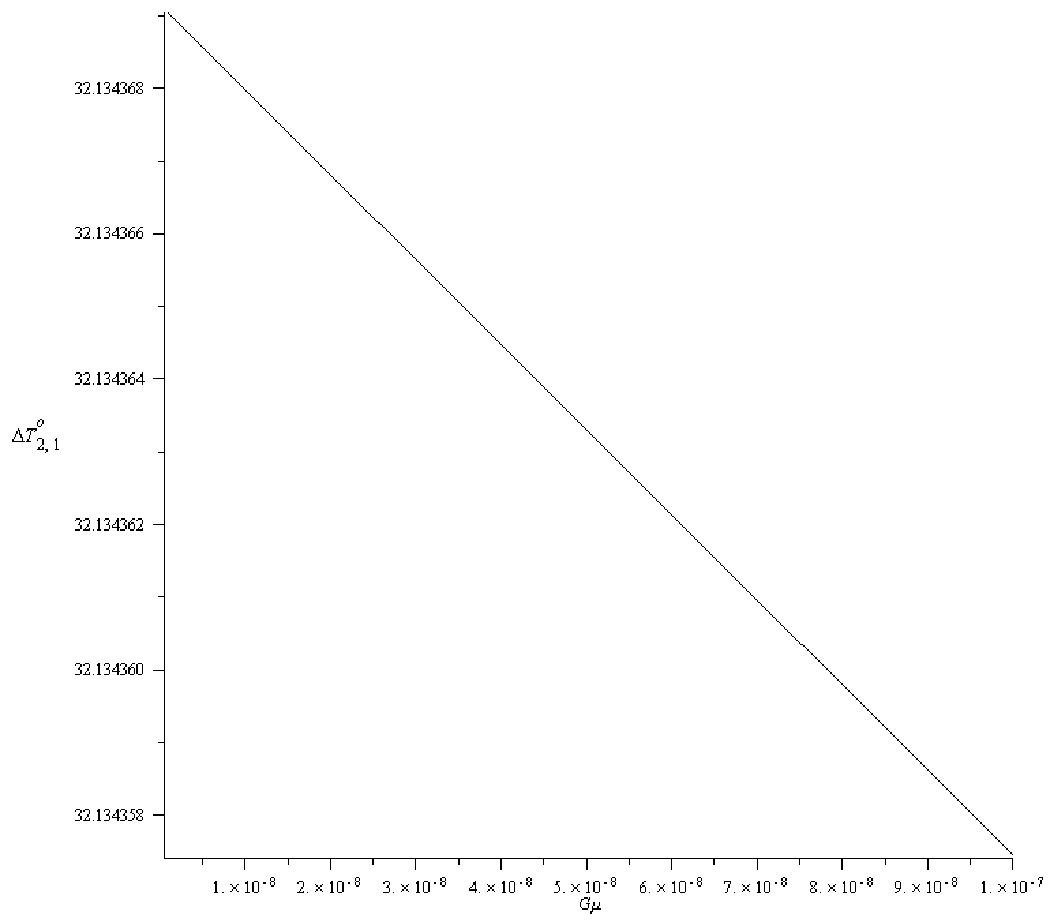}
  \caption{The dependence of time delay between two images on the opposite side on the parameter $G\mu$
  representing the tension of cosmic string piercing the Schwarzschild black hole.}
\end{figure}

\end{document}